# Prospects for Multiple Weak Gauge Boson Signals at Hadron Colliders

M. Y. Hussein

*Department of Physics, College of Science, University of Bahrain, Bahrain*

Abstract

We review the status of theoretical calculations relevant for electroweak physics at hadron colliders. The large parton flux at high energy gives rise to events where different pairs of partons interact contemporarily with large momentum exchange. We present calculations of the cross-section for two, three and four weak gauge bosons production in single and multiparton interactions.

## I. INTRODUCTION

The Standard Model, which has been successful in describing physics below the 100 GeV scale contains too many undetermined parameters to be considered as a fundamental theory, so it is interesting to search and test this theory deeply [1]. The potential studies of the electro-weak gauge boson production, which recognized along ago have provided important means to examine the Standard Model. Pair production of gauge bosons has been recognized at the high-energy hadron-hadron and electron-positron colliders. The total cross section for the production of triple gauge boson has been calculated [2]. As a consequence, we are therefore motivated to updated the calculation taking into account the improvements of the knowledge of the parton distribution function which has improved as more precision deep inelastic and other data have become available [3].

Motivated by the great success of Fermi-lab Tevatron and future LHC experiments, we carry out the calculations for pair, three and four electro-weak –gauge-boson production in the SM for the energy range 2-20 TeV. We have considered all combinations for the electro-weak-gauge boson processes for both single and double scattering mechanism:

$$\begin{aligned} pp &\to W^+W^-, W^+W^+, W^-W^-, ZZ \\ pp &\to W^+W^-W^+, W^+W^-Z, W^{\pm}ZZ, ZZZ \\ pp &\to W^+W^-W^+W^-, W^+W^-ZZ, ZZZZ \end{aligned} \quad (1)$$

In Section II, we present an overview of double parton scattering collision mechanism. Numerical results for multiple weak gauge boson pair production in section III. Section IV contains our conclusions.

## II. Double Parton Scattering Mechanism

Multiple parton interaction processes, where different pairs of partons interact contemporarily with large momentum exchange, become experimentally important at high energies because of the growing flux of partons. The importance of double parton scattering at the Large Hadron Collider (LHC) has been disscused [4].

The possibility of hadronic interactions with double parton scattering collisions was foreseen on rather general ground long ago [5]. The AFS collaboration at the CERN ISR has claimed evidence for the double scattering in pp collision [6]. The collider detector at FERMILAB CDF has observed the process by looking at final state with three mini-jets and one photon [7].

The multiple parton scattering occurs when two or more different pairs of parton scatter independently in the same hadronic collision. In principle double scattering probes correlation between partons in the hadron in the transverse plane, which provides extra additional information on hadron structure.

If the scattering event is characterized by high energy, then parton-parton correlation can be assumed to be negligible. With these assumptions the cross-section for a double collision leads in the case of two distinguishable parton interactions to the simplest factorized expression [5]

$$\sigma_{DS} = \frac{\sigma^a \sigma^b}{\sigma_{eff}} \tag{2}$$

Here $\sigma^a$ represents the single scattering cross section

$$\sigma_a = \sum_{i,j} \int dx_A dx_B f_i(x_A) f_j(x_B) \hat{\sigma}_{ij \to a} \tag{3}$$

With $f_i(x_a)$ is the standard parton distribution and $\hat{\sigma}_{ij \to a}$ represents the sub-process cross section. If the two interactions are distinguishable, double counting is avoided by replacing Eq. (2) by

$$\sigma_{DS} = \frac{\sigma^a \sigma^b}{2\sigma_{eff}} \tag{4}$$

Where $\sigma_{eff}$ is the effective cross section and it enters as a simple proportionality factor in the integrated inclusive cross section for a double parton scattering $\sigma_{DS}$. The value of $\sigma_{eff}$ represents therefore the whole output of the measure of the double parton scattering process, which on the other hand has shown to be in agreement with the available experimental evidence. The experimental value quoted by CDF [7] is $\sigma_{eff} = 14.5 \pm 1.7^{+1.7}_{-2.3} mb$. It is believed that $\sigma_{eff}$ is largely independent of the center-of-mass energy of the collision and on the nature of the partonic interactions.

Giving the potential importance of double parton scattering as a background to a new physics searches at the LHC, it is important to calibrate the effect by measuring $\sigma_{eff}$ using a known, well understood Standard Model process which suggests that, like-charged W pair production is much smaller than opposite-charged production, which suggests that it is the best place to look for additional double parton scattering contributions.

## III. Multiple Weak Gauge Boson Production

Electroweak measurments of multi-gauge boson production are very important part of the physics program of the Tevatron and the LHC. Both experimental collaboration at the Tevatron, CDF [7] and DO [8] has performed studies of pair search such as $pp \to W^+W^-, W^\pm Z, ZZ....,$. Some events has been found above the background, in accordance with Standard Model production. When the energy is going to increase at the upgraded Tevatron and at the Large Hadron Collider LHC, the precise measurements of the weak gauge boson production will be an important physics goal.

Hadronic multiple production of electro-weak vector boson received a lot of attention in the literature as they can used to measure the boson trilinear couplings predicted by the Standard Model [9]. The search for Higgs boson between 100-200 GeV and the determination of its properties, and the measurements of electroweak precision observables, is intimately connected to pair production of vector boson as also the production is expected to give a potential importance as a background to new physics searches at the LHC. The tree-level of $pp \to W^+W^-, W^\pm Z, ZZ....,$ as well as $pp \to W^+\gamma, Z\gamma,....,$ pairs were computed long ago [1].

The purpose of this section is to quantify the expected cross-section for weak boson production at different energies for both single and double scattering processes. We start our analysis by calculating the total cross–section for vector boson pair production.

Figure 1: shows the total cross section for $WW$ pair production in $pp$ as a function of collider energies. We use the MRST parton distribution from Ref. [10] and the most recent values for the electro weak parameters. The matrix elements are obtained using MADGRAPH [11]. In single parton scattering the cross section for like-charged and opposite-charged are differ by about two orders of magnitude. The cross section for opposite-charge ($W^+W^-$) production from double scattering contribution is smaller by two orders of magnitude than the single scattering process, while the cross sections for like-charged ($W^+W^+, W^-W^-$) production from double scattering contributing is smaller only factor (2.6-2) than the single scattering case. All results are in agreement with Ref. [12].

Figure 2: shows the vector boson pair $ZZ$ production cross-section as function of center-of-mass energy. The cross sections from double are smaller by order of magnitude than the single scattering case.

The comparison of the total number expected of $WW$ and $ZZ$ events expected for $L = 10^5 \, pb^{-1}$ at the LHC from single and double scattering, assuming $\sigma_{eff} = 14.5 mb$ is shown in Table 1.

We carry out the calculations for three electro-weak gauge boson production in the SM, relevant to future experiments at a collider upgrade for luminosity and center of mass energy [13,14]. We have consider all the combinations for the three electro-weak gauge boson processes:

$$pp \to W^+W^-W^\pm, W^+W^-Z, W^\pm ZZ, ZZZ \qquad (5)$$

Figure 3: represents the total cross section for processes with three-gauge boson production for single and double parton scattering. The cross rates at $\sqrt{s} = 14 TeV$ are about $(10^{-1} pb - 10^{-5} pb)$ for $ZZZ$ process (Lowest) and are about $(7 \times 10^{-1} pb - 2 \times 10^{-3} pb)$ for $W^+W^-W^\pm$ process (Highest) in single parton scattering as given in Ref. [15] and double parton scattering. Although the cross sections are rather small but they can get sizable number of events if the luminosity is sufficiently high.

Table 2: summarizes the number of expected events in various three vector gauge boson channels (recall these are leading-order estimate only, with no branching ratios), assuming $\sigma_{eff} = 14.5 mb$. however, the absolute event rates shown in Table 2 are sensitive to overall measured and theoretical uncertainties.

Four weak boson production $pp \to W^+W^-W^+W^-, W^+W^-ZZ, ZZZZ$ are evaluated in Ref. [15]. The direct four boson contributions are small compared to other signals; the $4W$ and $4Z$ cross-section at LHC are $\sigma(4W) = 2.2 \times 10^{-3} pb$ and $\sigma(4Z) = 2 \times 10^{-5} pb$ as shown in Ref. [16]. We calculated the total cross section for the production of four bosons using double parton scattering. The set of integrated processes is:

$$(qq)_1 + (qq)_2 \to WWWW, WWZZ, ZZZZ \qquad (6)$$

Figure 4: shows the total cross section in $pp$ collisions as a function of the collider energy. The double parton scattering is giving very small values for cross section and the expected numbers of events per year is less than one, so these processes are not (expected to be) observable and probably beyond experimental capability. However, it still shows a phenomenon, which can be expected at high energies at upgraded Tevatron.

Table 3: summarizes the number of expected events in various four vector gauge boson channels (recall these are leading-order estimate only, with no branching ratios), assuming $\sigma_{eff} = 14.5 mb.$.

**IV. Conclusions**

Electroweak measurements are a very important part of the physics program of the Tevatron and the LHC. Accurate predications are needed to fully utilize the potential of colliders for electroweak measurements. Of particular interest are the search for the Higgs boson and the determination of its properties, and the measurement of the electroweak precision observable.

The theoretical predications for weak boson and Higgs boson production have become increasingly accurate over the past few years. However, there is still much to do. In particular a calculation of cross sections for multiple gauge boson production in the SM. These productions processes must be quantified for future experiments at the upgraded Tevatron in order to examine the electroweak Standard Model, and to explore physics beyond Standard Model. Furthermore, since heavy particles often decay to gauge bosons as a final state, it is important to examine the multi-gauge boson production in searching for new physics. However, in the hadron collider environment, the large QCD backgrounds may cause the observation impossible for those events if one or more of the gauge bosons decay hadronically. Individual channels with hadronic decays should be studied on a case by case.

We have shown that multiple weak gauge boson production provides a relatively a way of searching for, and calibrating, double parton scattering at the LHC.

TABLES

Table 1: The expected number of VV events expected for luminosity = $10^5$ $pb^{-1}$ at the LHC from single and double scattering, assuming $\sigma_{eff} = 14.5 mb$.

|  | $N(W^+W^-)$ | $N(W^+W^+)$ | $N(W^-W^-)$ | $N(ZZ)$ |
|---|---|---|---|---|
| Single scattering | 80,000,000 | 80,000 | 25,000 | 1,300,000 |
| double scattering | 42,000 | 26,00 | 15,000 | 30,000 |

Table 2: The expected number of VVV events expected for luminosity = $10^5$ $pb^{-1}$ at the LHC from single and double scattering, assuming $\sigma_{eff} = 14.5 mb$.

|  | $N(W^+W^-W^\pm)$ | $N(W^+W^+Z)$ | $N(W^\pm ZZ)$ | $N(ZZZ)$ |
|---|---|---|---|---|
| Single scattering | 70,000 | 42,000 | 12,000 | 10,000 |
| double scattering | 210 | 110 | 9 | 〈1 |

Table 3: The expected number of VVVV events expected for luminosity = $10^5$ $pb^{-1}$ at the LHC from double scattering, assuming $\sigma_{eff} = 14.5 mb$.

|  | $N(W^+W^-W^+W^-)$ | $N(W^+W^+ZZ)$ | $N(ZZZZ)$ |
|---|---|---|---|
| Single scattering | 250 | 25 | 2.5 |
| double scattering | 〈1 | 〈1 | 〈1 |

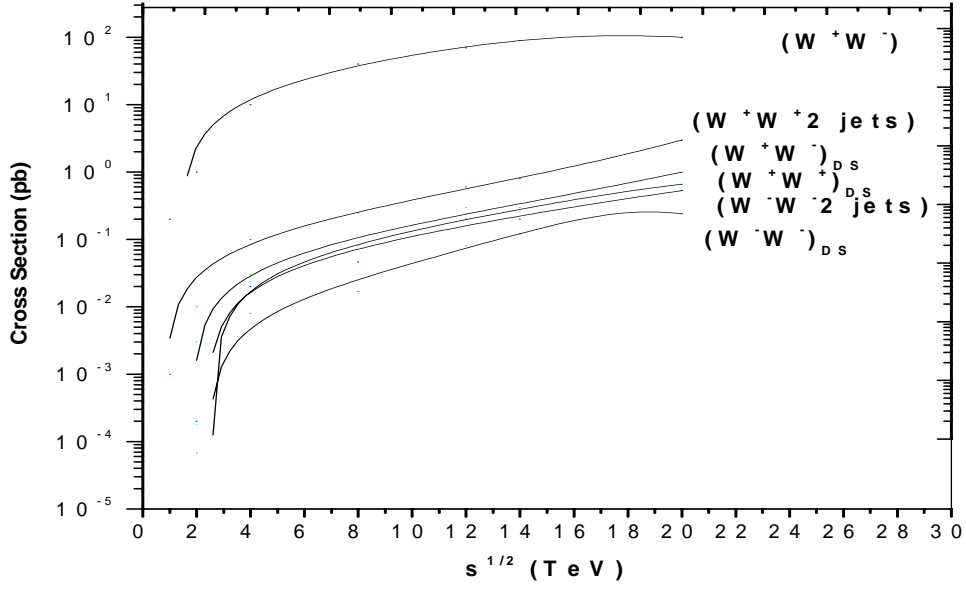

Figure 1: Total cross section for W pair production versus the c.m. energy $\sqrt{s}$ in $pp$ collision corresponding to single and double parton scattering assuming $\sigma_{eff} = 14.5 mb$.

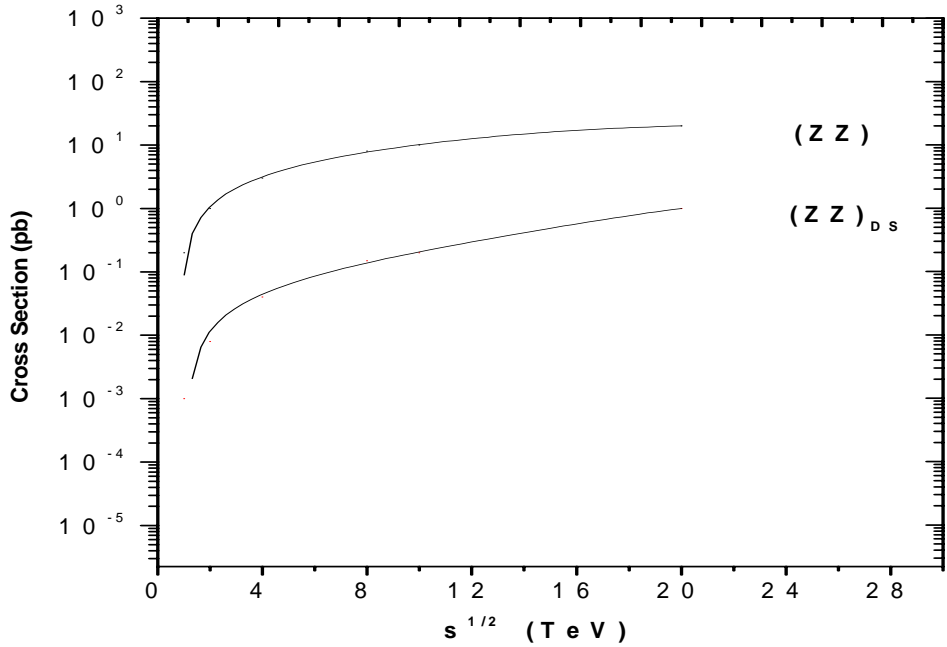

Figure 2: Total cross section for Z pair production versus the c.m. energy $\sqrt{s}$ in $pp$ collision corresponding to single and double parton scattering assuming $\sigma_{eff} = 14.5 mb$.

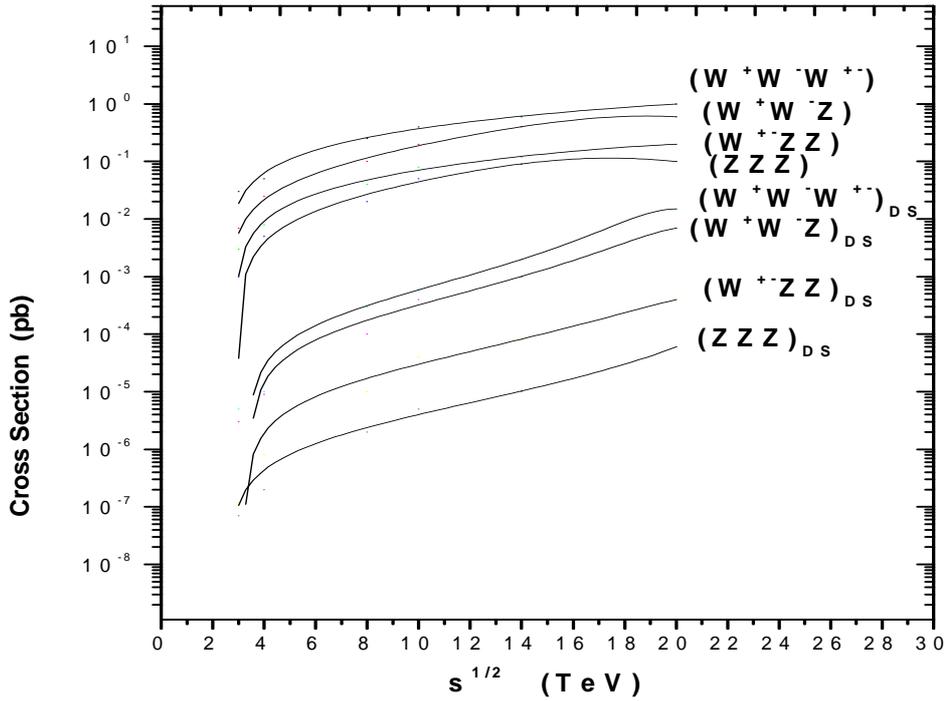

Figure 3: Total cross section for triple weak gauge production versus the c.m. energy $\sqrt{s}$ in $pp$ collision corresponding to single and double parton scattering assuming $\sigma_{eff}=14.5 mb$.

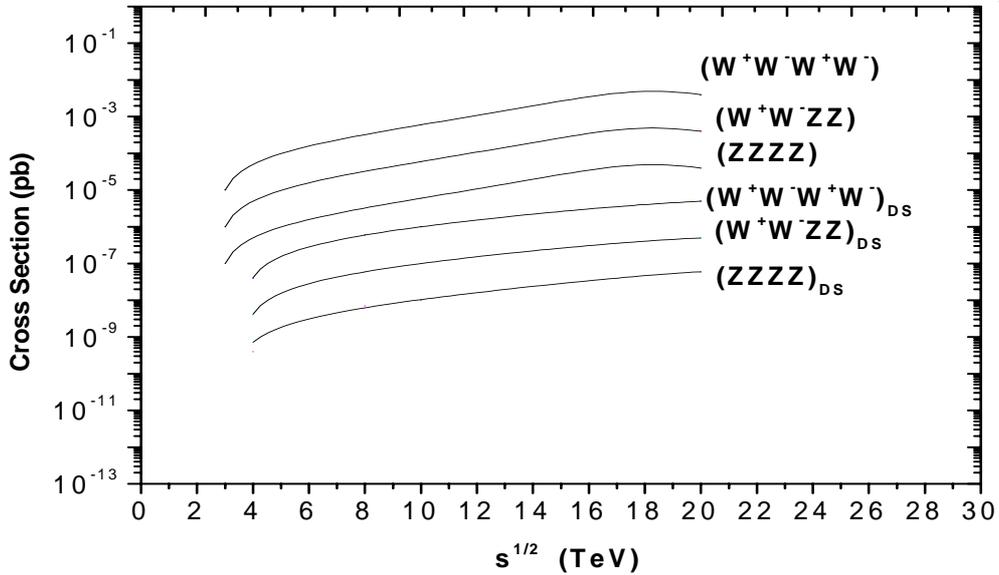

Figure 4: Total cross section for four weak gauge production versus the c.m. energy $\sqrt{s}$ in $p\bar{p}$ collision corresponding to single and double parton scattering assuming $\sigma_{eff}=14.5 mb$.

**Acknowledgments**

I am grateful for conversations and correspondence with Prof. W. J. Stirling.